\documentclass[amsmath,amssymb,pra,aps,twocolumn,mathbbm,superscriptaddress]{revtex4}
\usepackage{amssymb}
\usepackage{graphicx}
\usepackage{times}
\usepackage{subfigure}

\begin{document}
\title{Simulating superluminal Physics with Superconducting Circuit Technology}
\author{Carlos Sab{\'\i}n}
\affiliation{Instituto de F{\'i}sica Fundamental, CSIC,
Serrano, 113-bis,
28006 Madrid, Spain}
\email{Correspondence to: csl@iff.csic.es}
\author{Borja Peropadre}
\affiliation{Department of Chemistry and Chemical Biology, Harvard University, Cambridge, Massachusetts 02138, United States}
\affiliation{Quantum Information Processing group, Raytheon BBN Technologies, 10 Moulton Street, Cambridge, Massachusetts 02138, USA}
\author{Lucas Lamata}
\affiliation{Department of Physical Chemistry, University of the Basque Country UPV/EHU, Apartado 644, E-48080 Bilbao, Spain}
\author{Enrique Solano}
\affiliation{Department of Physical Chemistry, University of the Basque Country UPV/EHU, Apartado 644, E-48080 Bilbao, Spain}
\affiliation{IKERBASQUE, Basque Foundation for Science, Maria Diaz de Haro 3, E-48013 Bilbao, Spain}

\begin{abstract}
We provide tools for the quantum simulation of superluminal motion with superconducting circuits. We show that it is possible to simulate the motion of a superconducting qubit at constant velocities that exceed the speed of light in the electromagnetic medium and the subsequent emission of Ginzburg radiation. We consider as well possible setups for simulating the superluminal motion of a mirror, finding a link with the superradiant phase transition of the Dicke model.
\end{abstract}
\maketitle
\section{Introduction}
The fact that physical signals containing energy or information are not allowed to travel faster than the speed of light in vacuum is one of the best established facts in modern physics, and the cornerstone of one of its deepest principles, namely, causality. However, this does not prevent one from considering interesting features of the considered model at superluminal speeds. On the one hand, not all velocities are physical, in the sense that they do not need to carry any content of information. A related example can be found in Ref.~\cite{french}. On the other hand, experiments  do not typically take place in vacuum, but in some medium, in which light moves at slower speeds. Therefore, it may be possible to consider physical motion at velocities exceeding those of light in the medium but not in vacuum. Classically, this gives rise to the well-known Cerenkov effect, where a moving electric charge generates classical electromagnetic radiation. In the quantum realm, the counterpart of the Cerenkov effect involves a neutral body or any sort of perturbation moving at superluminal speeds, i.e., the so-called Ginzburg radiation \cite{ginzburg, quantumcerenkov,ginzburgbec,faccio}. 

Quantum simulators are controllable quantum platforms aiming at reproducing the properties of complex quantum systems. They will soon be able to outperform classical computers  and reach quantum supremacy. Quantum simulators can also be conceived as helpful tools which enhance our understanding of modern theoretical physics by allowing us to go beyond its fundamental laws. Along this vein, phenomena and effects which are not amenable to experiments due to technical or fundamental reasons, are now within reach of the burgeoning field of quantum simulations, ranging from magnetic monopoles~\cite{monopole} to traversable wormholes~\cite{qswh} or tachyons ~\cite{tachyons}.

Superconducting circuits are one of the most promising quantum platforms for the development of scalable quantum technology and could be among the first in demonstrating quantum supremacy \cite{preskill, boixo}. In parallel, they are also a natural testbed for relativistic physics in quantum mechanics and quantum field theory, either in direct or simulated observations. For instance, the generation of photons out of the vacuum due to motion of mirror-like boundary conditions at relativistic speeds, namely, the dynamical Casimir effect (DCE), has been demonstrated in superconducting circuit architectures \cite{casimirwilson}. Along the same lines, a quantum simulation of the generation of acceleration radiation by means of relativistic motion of a superconducting qubit has been proposed \cite{simoneunruh}, i.e., cavity-enhanced Unruh effect. While the ultrafast variation of magnetic fluxes allows to achieve highly relativistic effective velocities, exploring both DCE and Unruh physics, breaking the light barrier with superconducting circuits, either in a medium or a quantum simulation, remains unexplored. Indeed, DCE experiments are restricted to velocities well below this threshold.

In this Letter, we provide tools for quantum simuation of superluminal motion with superconducting circuits. By superluminal we mean both exceeding the velocity of light in the medium -which is in principle possible as a real effect- and in vacuum -which can be only conceived in a simulator. We show that it is possible to simulate with current platforms the motion of a superconducting qubit at constant speeds exceeding the speed of light, even in vacuum, in the electromagnetic environment provided by a transmission line resonator. Remarkably, this effective superluminal motion can trigger the emission of Ginzburg radiation. We discuss as well the possibility of achieving superluminal constant velocities in the simulation of mirror-like boundary motion. We find that a setup similar to the one required for testing the Dicke model in the thermodynamic limit can be used for the simulation of the Hamiltonian of a mirror moving at superluminal speeds. Moreover, we find a link to Dicke superradiant phase transition. Notice that the emission of radiation by means of superluminal motion has two key differences with Unruh and DCE physics, namely that it only appears above the threshold of the speed of light in the medium and that it does not require accelerations. Moreover, it is worth highlighting that previous proposals for simulating relativistic motion with superconducting circuits were fundamentally constrained to the subluminal regime. Therefore we develop here a new set of ideas, both from the conceptual and the theoretical side. Note also that we are considering the superluminal motion of a superconducting qubit and a mirror, which is different from superluminal propagation of the microwave radiation itself in superconducting circuits \cite{mahmoudi}.

\section{Ginzburg radiation}
Let us start by showing how a qubit interacting with a single resonator mode via a quantum Rabi Hamiltonian emits radiation when moving at superluminal speeds --a particular case of Ginzburg radiation.
This model is described by the Hamiltonian
\begin{equation}
\label{physHam}
\mathcal{H} = \omega_0 a^\dagger a + \frac{\omega_q}{2}\sigma_z + \mathcal{H}_I(x_q) ,
\end{equation}
where $\omega_q$ is the qubit energy spacing, $\sigma_z$ and $\sigma_x$ are the usual Pauli operators acting on the qubit Hilbert space, and $\hbar = 1$. We assume that the system dynamics effectively involves a single resonator mode, described by annihilation and  creation operators $a$ and $a^\dagger$, respectively, of frequency $\omega_0= c k$ and wave vector $k = \pi/L$. Here, $L$ is the resonator length and  $c$ is the speed of light, which in the case of a superconducting resonator takes a typical value $c_0/3$, where $c_0$ is the speed of light in vacuum.  

The interaction Hamiltonian is
\begin{equation}
\label{physHI}
\mathcal{H}_I(x_q) =  g \cos{\left( k x_q \right)} \sigma_x (a^\dagger + a),
\end{equation}
where $g$ is the coupling strength and $x_q$ the qubit position~\cite{Shanks2013}. This model is the standard cavity Quantum Electrodynamics approximation to the full Quantum Field Theory matter-radiation interaction Hamiltonian, and has been used in the literature to describe the emission of radiation by an atom moving at relativistic speeds \cite{scully,hureply,scullycomment,scullytheory}. 

Within perturbation theory, the probability of photon emission of a qubit starting in the ground state, with the field starting in the ground state as well, reads
\begin{equation}
P_e=g^2\left|\int^T_0\,dt\,e^{i(\omega_q+\omega_0)\,t}\cos{k\,x_q(t)}\right|^2,
\end{equation} 
where  $x_q(t)$ is possibly time-dependent. If the qubit is static this non-RWA probability is eventually negligible. However, if the qubit moves at constant velocity
\begin{equation}\label{Eq1Velocity}
x_q(t)=x_0+v\,t
\end{equation}
and assuming for simplicity $x_0=0$ we find:
\begin{equation}
P_e=g^2\left|\int^T_0\,dt\,e^{i(\omega_q+\omega_0)\,t}\cos{k\,v t}\right|^2.
\end{equation}
Therefore, if the velocity is:
\begin{equation}
v=\frac{\omega_q+\omega_0}{\omega_0}c,
\end{equation} 
the probability of photon emission is resonantly enhanced. Notice that this activation of the counterrotating terms of the Hamiltonian resembles the cavity-enhanced Unruh effect \cite{scully,hureply,scullycomment,scullytheory}, a similarity first noted by Ginzburg in a more general context~\cite{ginzburg}. However, in this case there is no acceleration, and the effect comes from the superluminality of the qubit motion.

Let us highlight further analogies between the acceleration radiation scenario and the current constant-velocity one. If we consider an oscillatory motion starting at the center of the cavity and oscillating with frequency $\omega$ along the full resonator length \cite{simoneunruh} $x_q(t)=L/2 +L/2\cos{\omega\,t}$, then we have $k x_q(t)=\pi/2 +\pi/2\cos{\omega\,t}$ and $\cos{k x_q(t)}=-2\sum_{0}^{\infty}(-1)^k J_{2k+1}(\pi/2)\cos{(2k+1)\omega t}$, where the $J_{2k+1}$'s are Bessel functions of the first kind. Since $J_1(\pi/2)>>J_3(\pi/2)$, we can finally write
\begin{equation}
\cos{k\,x_q(t)}\simeq-2J_1(\pi/2)\cos{\omega\,t},\label{eq:bess}
\end{equation}
where $J_1(\pi/2)$ is the value of the Bessel function of the first kind evaluated at $\pi/2$. Using Eqs. (\ref{physHI}) and (\ref{eq:bess}), we find that the interaction Hamiltonian of this oscillatory motion would be
\begin{equation}
\label{physHI2}
\mathcal{H}_I(x_q) \simeq -2 g J_1(\pi/2) \cos{\left( \omega\,t \right)} \sigma_x (a^\dagger + a),
\end{equation}
which would be, as seen in Eq.~(\ref{Eq1Velocity}), the same as the interaction Hamiltonian in the case of a trajectory with constant velocity $\frac{\omega}{\omega_0}\,c$ starting at $x=0$ with a coupling strength $-2gJ_1(\pi/2)$. Therefore, we conclude that we can approximate a motion with constant velocity along the resonator by an oscillatory motion which starts at the center of the resonator and spans from mirror to mirror.

Putting all the above together, we find that it is possible to simulate superluminal constant velocities using existing experimental techniques. In the circuit QED architecture proposed in Ref.~\cite{simoneunruh}, the interaction Hamiltonian has the following dependence on the external magnetic flux,
\begin{equation}\label{physHI3}
\mathcal{H}_I(f) =  g_0 \cos{\left(  f \right)} \sigma_x (a^\dagger + a).
\end{equation}
Here, $f=\phi/\phi_0$ is the magnetic frustration, where $\phi$ and $\phi_0$ are the magnetic flux and  flux quantum, respectively.
Identifying the flux as
\begin{equation}\label{eq:fprof}
f= k\,x_q,
\end{equation}
the Hamiltonians in Eqs.~\eqref{physHI} and~\eqref{physHI3} are equivalent. Therefore, the modulation of the effective coupling constant mimics the motion of the qubit $x_q(t)$ inside the transmission-line resonator (TLR).  In the case described above, this means that 
\begin{equation}\label{eq:simf}
f=\frac{\pi}{2}+\frac{\pi}{2}\cos{(\omega_q+\omega_0)t}
\end{equation}
implements an oscillatory motion around the center of resonator with frequency $\omega_q+\omega_0$, which would be equivalent to a motion with constant superluminal velocity, as shown above.

In the resonant case $\omega_q=\omega_0$, we will have an effective velocity $v=2c$, which for typical circuit QED architectures would still be below $c_0$. Adding a large detuning would enable the simulation of velocities that go even beyond $c_0$. 

\begin{figure}[h!]
\includegraphics[width=0.5\textwidth]{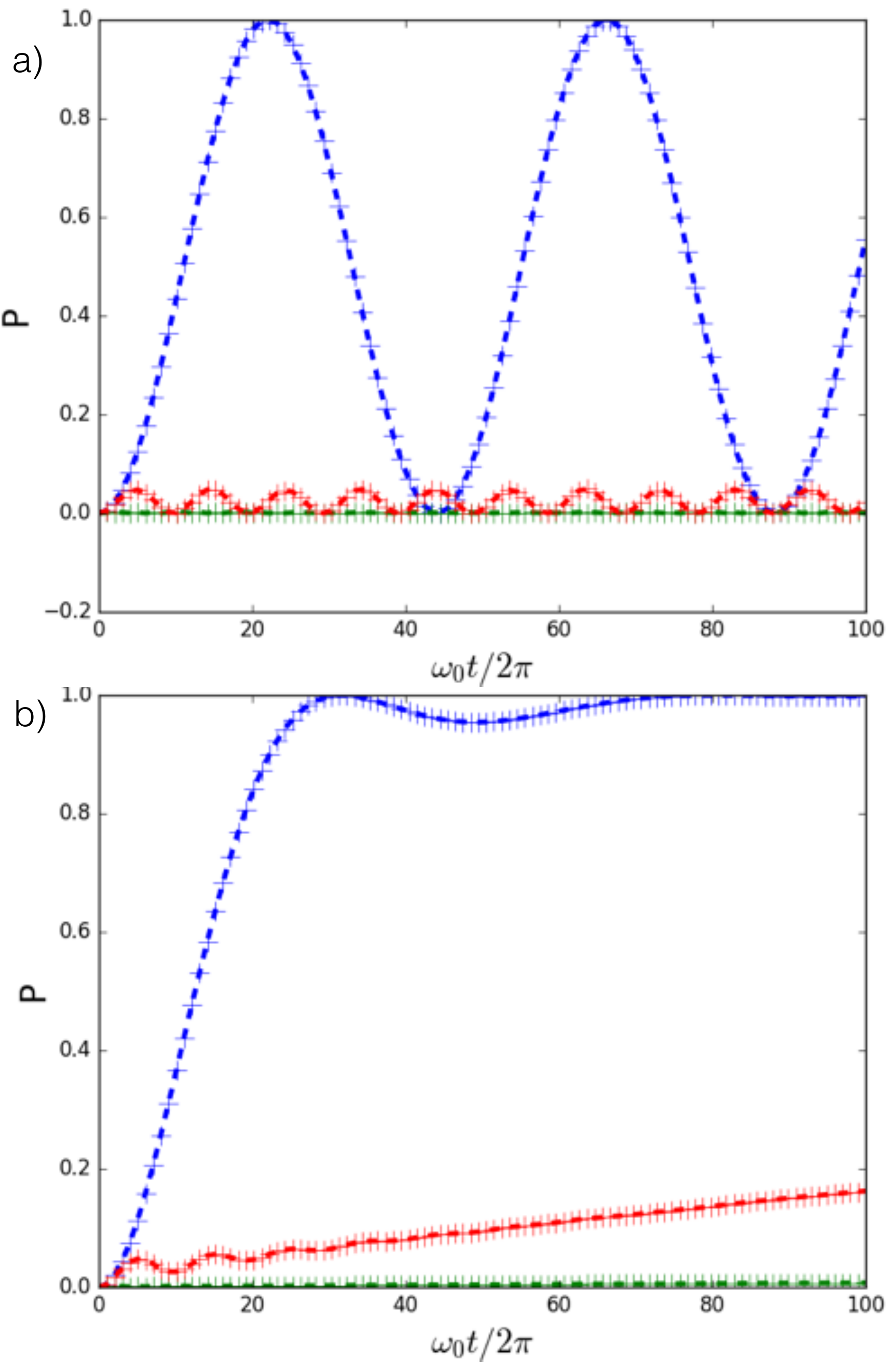}
\caption{(Color online) Probability of excitation for a qubit which is initially in the ground state and follows trajectories $x_q(t)=\frac{\pi}{2}+\frac{\pi}{2}\cos{\omega\,t}$ (crosses) and $x_q(t)=\frac{\omega}{\omega_0}c\,t$ (dashed curves), with $g=0.02$ for the dotted curves and $g=-2\,J_1(\pi/2)0.02$ for the dashed ones. The curves always superpose, showing the excellent accuracy of the approximation in Eq. (\ref{eq:bess}). The frequencies are $\omega=2\omega_0$ and $\omega_q=\omega_0/2$ (green), $\omega_q=0.9\omega_0$ (red) and $\omega_q=\omega_0$ (blue). Therefore, in all cases $v=2\,c$, but only the blue curves represent the resonance $\omega=\omega_q+\omega_0$. The qubit decay parameters are $\Gamma = 0.002$, and $T_2/T_1 = 0.67$, and we consider two cavity decay rates: a) $\kappa= 0.001$, and b) $\kappa= 0.1$ (bad-cavity limit), in units of $\omega$.}
\label{fig1}
\end{figure}

In Fig.~\ref{fig1}, we plot the results of numerical simulations. The dynamics is governed by a master equation where we introduce a cavity decay rate $\kappa$, a decay parameter $\Gamma$ accounting for dissipative processes, as well as a decay $\Gamma_{\varphi}$ for the dephasing of the qubits. The energy relaxation time and phase coherence time are denoted with $T_1 = 1/\Gamma$ and $T_2 = 1 / \Gamma_{\varphi}$, respectively. We consider typical parameters in current experiments~\cite{Zhang2016}, $\Gamma/\omega=10^{-3}$, $\Gamma_{\varphi}/\omega=5\cdot10^{-4}$. We see the excellent accuracy of the approximation in Eq. (\ref{eq:bess}) and the neat resonance for the velocity $(\omega_q+\omega_0)/\omega_0$. In Fig. \ref{fig1}a the decay rate of the cavity is small and thus we observe perfect Rabi oscillations as expected from anti-Jaynes-Cummings dynamics \cite{simoneunruh}. In Fig.~\ref{fig1}b, the decay rate is much larger, entering into the bad-cavity limit \cite{diquewall}. In this case, the oscillations are washed out and the qubit is projected onto its excited state~\cite{laura}. To retrieve the qubit states, one may use auxiliary resonators with dispersive microwave drivings to perform projective measurements of the qubits in the computational basis~\cite{ReviewGoran}. 

\section{A mirror moving at superluminal speeds}
Now we will consider a different scenario, where a mirror moves at superluminal speeds. It has  been shown  that the motion of optical boundaries \cite{guerreiro1, guerreiro2} or the perturbation of the refractive index \cite{faccio}  at constant and superluminal speeds generates photons out of the vacuum. This phenomenon somehow resembles the DCE, but it is radically different: there is no acceleration and it only appears at superluminal speeds. Moreover, it is also different from the Cerenkov effect, which requires the presence of a charge and is classical.

Although the DCE with oscillating motion is the most conspicuous example, other instances of boundary motion have been considered in the literature  \cite{vesnitskii1, vesnitskii2,sarkar,dodonov}. However, the case of a mirror moving at superluminal speeds remains unexplored.

The DCE was observed in an open microwave coplanar
waveguide interrupted by a single superconducting quantum interference device (SQUID) operated well below its plasma frequency \cite{casimirwilson}. Under the latter condition, the SQUID implements an effective mirror-like boundary condition for the superconducting flux field, which can be described by a standard quantum 1D bosonic field. The effective position of the boundary condition depends on the particular value of the magnetic flux threading the SQUID, thus its ultrafast variation amounts to motion of the mirror at relativistic speeds, which generates a two-mode squeezing operation on the field propagating along the transmission line. The DCE can be produced as well for different boundary conditions, such as the ones of a superconducting resonator interrupted by one \cite{simoen} or two SQUIDs \cite{idathesis}. In general, it will appear in a cavity with time-dependent length, where the variation of the length takes place at relativistic speeds.

We consider now a 1D cavity of time-dependent length. In particular, let us assume that the cavity has a fixed length $L$ until $t=0$ and then the length changes in time, $L(t)$. The effective Hamiltonian for this system has been derived in Refs.~\cite{razavy1,razavy2}:
\begin{eqnarray}\label{eq:hamsuperl}
H_{\rm eff}&=&\sum_{n}\omega_n(t) \left(a_n^\dagger a_n+\frac{1}{2}\right)+
\frac{\dot L (t)}{L(t)}\sum_{n}\sum_{j\neq n}\\& &\!\!\!\!(-1)^{n+j}\frac{jn}{j^2-n^2}\sqrt{\frac{n}{j}} (a_n^\dagger a_j^\dagger+ a_n^\dagger a_j-a_n a_j^\dagger - a_n a_j),\nonumber
\end{eqnarray}
where
\begin{equation}
\omega_n(t)=\frac{\pi c n}{L(t)},
\end{equation}
and $\dot L(t)$ is the time derivative of $L(t)$.

In the DCE implementation a constant DC flux field is modulated through a small harmonic AC field of frequency $\omega_d$. This results in an effective harmonic motion of the mirror characterized by a small oscillation amplitude. Considering $L(t)=L(1+\delta\sin{\omega_d t})$ with $\delta\ll 1$, it is straightforward to realize that $\frac{\dot L(t)}{L(t)}\simeq v_{max}\cos{\omega_d t}$, which in the interaction picture leads to two-mode squeezing proportional to $v_{max} $ if $\omega_d=\omega_k+\omega_j$. Therefore, the DCE is a particular case of the model embodied by Eq. (\ref{eq:hamsuperl}).

However, the achievable mirror velocities in the celebrated circuit Quantum Electrodynamics (circuit QED) implementation of the DCE are severely limited \cite{casimirwilson}. In particular, the maximum velocity of the harmonic motion is $v_{max}=\delta L_{\rm eff}\,\omega_d$
where $\delta L_{\rm eff}$ is the amplitude of the oscillation.  $\omega_d$ needs to be well below the SQUID plasma frequency which typically means $\omega_d<20\,\operatorname{GHz}$ --it was  $10\, \operatorname{GHz}$ in Ref.~\cite{casimirwilson}.
Moreover, the  SQUID-mirror equivalence only works if 
$k_{\omega}\,L_{\rm eff}\ll 1$, namely $L_{\rm eff}$ must be smaller than the relevant wavelengths.
Putting everything together it turns out that 
$v_{max} \ll 2\,c.$
Therefore, it is not possible to achieve the superluminal regime with the setup of Ref. \cite{casimirwilson}.

Now, let us consider $L(t)=L-v\,t$. Note that even if $v<c$, this trajectory is unphysical, since it predicts an infinite acceleration at $t=0$. Of course, this is not a concern in a simulated scenario. Using Eq. (\ref{eq:hamsuperl}),  we see that we have both two-mode squeezing and mode mixing proportional to 
\begin{equation}
\frac{ \dot L(t)}{L(t)}=-\frac{v}{L-v t}.
\end{equation} 
Note the obvious restriction $v t<L$, i.e. $L(t)>0$. We can consider that this is a restriction on time, not on velocity, and thus nothing prevents us from considering superluminal simulated velocities $v>c$. We can even restrict ourselves to shorter simulated times $v t\ll L$ where
\begin{equation}
\frac{\dot L(t)}{L(t)}=-\frac{v}{L}.
\end{equation} 
Under this approximation, the Hamiltonian becomes time-independent. Note that $c/L$ is the characteristic frequency scale of the system, so if we want velocities around $c$, the aim is to generate an interaction between the modes with a strength comparable to their frequencies, namely ultrastrong coupling among bosonic modes. 

More specifically, let us restrict the Hamiltonian in Eq. (\ref{eq:hamsuperl}) to the pair of lowest modes of a resonator, with frequencies $\omega_1$, $\omega_2$ where $\omega_2=2\omega_1=2\pi c/L$. What we obtain is a model of two coupled bosonic modes, with an interaction that depends on the effective velocity. Therefore, we propose to use a model of two coupled bosonic modes in order to simulate mirror motion at constant speeds, where the interaction strength $\Omega$ codifies the simualated velocity. In particular, the coupling strength of the squeezing part of the Hamiltonian is
\begin{equation}
\Omega=\frac{\sqrt{2}}{3}\frac{v}{L}\label{eq:squeezing2m}.
\end{equation}
Thus, for $v=c$ we find $\Omega/\omega_1\simeq0.15$.  Achieving this coupling strength and higher values in order to explore the superluminal region seems extremely challenging in a coupled-cavity setup, although it might be within reach in the case of SQUID-mediated coupling \cite{peropadre, peropadre2}. 

Alternatively, we can also simulate one of the modes with an array of $N$ qubits which are coupled to a single resonator mode. This is the Dicke model \cite{dique}, which is well-known in quantum optics and has also been studied, both theoretically and experimentally in circuit QED. The interest in the Dicke model comes chiefly from the appearance of a so-called superradiant phase transition, where spontaneous emission of the atoms is strongly enhanced as a result of collective quantum effects \cite{dique}.The actual existence and implications of this phase transition has been the subject of intense debate, both in cavity QED and circuit QED \cite{ciuti, oliver}. It will only take place if the physical Hamitonian is the actual Dicke hamiltonian, namely two coupled bosonic modes with no influence of extra terms, like  the so called diamagnetic term accounting for self-interaction --which is familiar in quantum optics.
In the Dicke model in the thermodynamic limit $N\gg1$ the qubits are represented as well by a single collective bosonic mode, and the coupling between this effective mode and the resonator mode is proportional to the number of qubits. In this way, we can take advantage of the enhancement of the intermodal coupling $\Omega\propto\sqrt{N}$ \cite{ciuti,fink}. Indeed, the celebrated superradiant phase transition would take place at a critical value of the coupling $\Omega_c=\sqrt{\omega_1\omega_2}/2=\pi\,c/(\sqrt{2}L)$. This would correspond to a superluminal velocity $v=3 c /2 \pi$. Note that we are assuming that the Hamiltonian is just the one of two ultrastrongly coupled bosonic modes, that is  we can neglect the diamagnetic term and any other extra terms \cite{ciuti,ripperlib}, which seems possible in some (e.g., Ref.~\cite{nakamura}) but not all (e.g., Refs.~\cite{oliver, rable}) superconducting circuit architectures. In this way, we find a remarkable analogy between the Dicke superradiant phase transition and the physics of a mirror moving at relativistic speeds, which can be seen as an additional motivation for an experimental test of the Dicke model in the $N\gg1$ limit with superconducting circuit technology. Notice that increasing the number $N$ of qubits amounts to increasing the square of the simulated mirror velocity $v^2$. So far, an array of 20 flux qubits coupled to a single resonator mode has been implemented in the laboratory, and the $\sqrt{N}$ enhancement of the coupling has been proved up to 8 qubits \cite{macha}.

\begin{figure}[h!]
\includegraphics[width=0.5\textwidth]{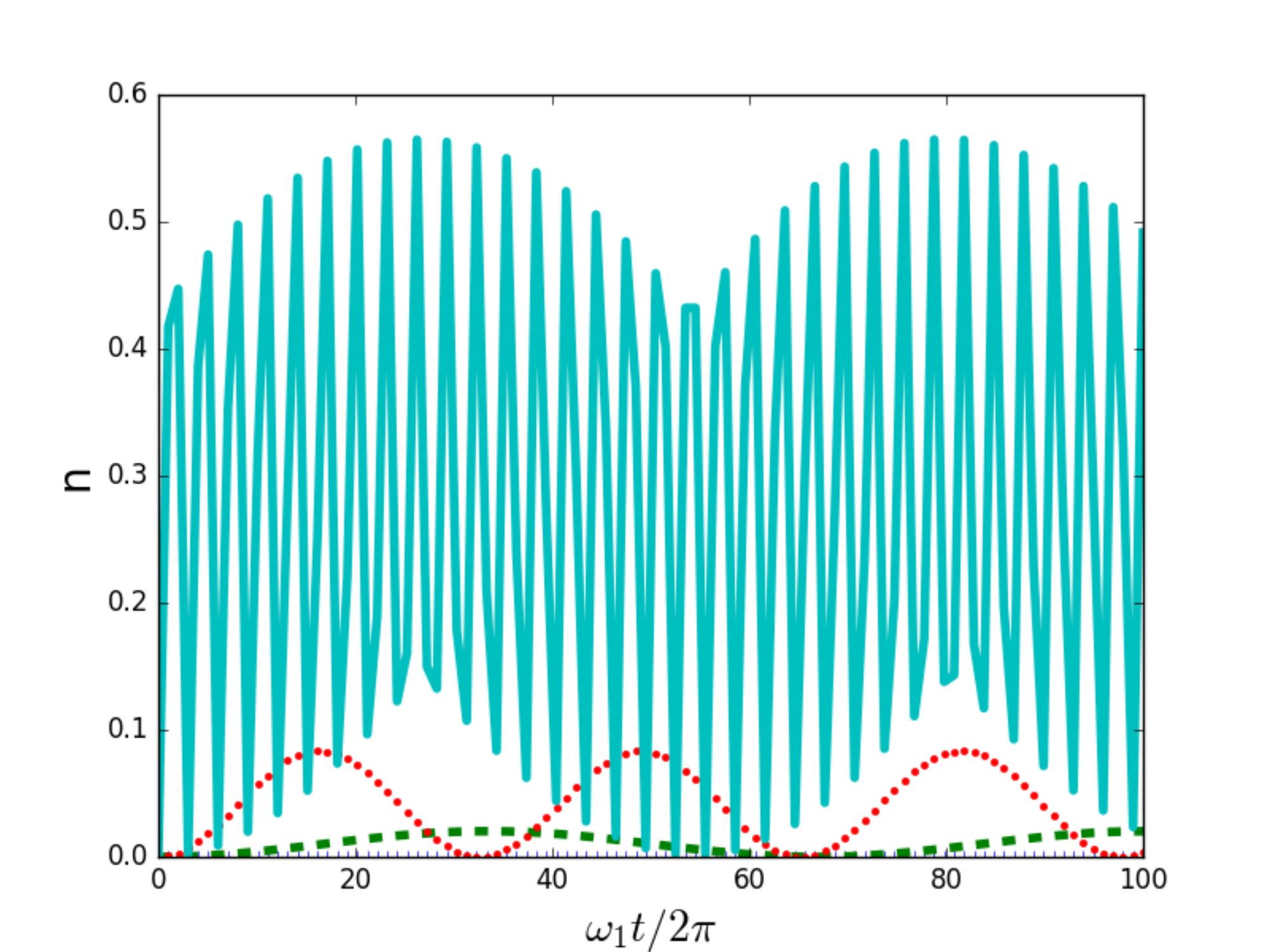}
\caption{(Color online) Total number of photons generated in the modes of frequencies $\omega_1$ and $\omega_2$ for a squeezing strength $\Omega$ given by Eq. (\ref{eq:squeezing2m}) and $v/c= 0.1$ (dark blue, crosses), $1$ (green, dashed), $2$ (red, dotted) and $3\pi/2$ (light blue, solid). The latter corresponds to the critical value of the analogue superradiant phase transition. Note that $\Omega/\omega_1\simeq 0.15\,v/c$. We consider a decay rate $\kappa = 0.001$.}
\label{fig2}
\end{figure}

In Fig.~\ref{fig2}, we plot numerical simulations of the dynamics of the two-mode model described above, including a decay rate $\kappa$. Starting from an initial vacuum, we observe generation of photons for simulated superluminal velocities, well above the average number of thermal photons at the $10-100\,\operatorname{mK}$ relevant regime of temperatures.  
To measure the number of photons in an implementation with superconducting circuits, one may employ standard circuit quantum electrodynamics techniques, e.g., dual path techniques~\cite{DualPathRoberto}. 

\section{Conclusions} 

We have provided tools for the quantum simulation of superluminal motion with state-of-the-art superconducting quantum technology. We have shown that the achievement of simulated velocities exceeding the speed of light in the electromagnetic environment, and possibly in vacuum, can be related to Unruh and DCE physics and, more surprisingly, to the superradiant phase transition of the Dicke model. Our results do not only open a new front in quantum simulations with superconducting
circuit technology, but also establish a natural arena for the analysis of phenomena such as
Casimir forces or quantum friction induced by Ginzburg radiation \cite{ginzburgbec}. Instead of using an
analog quasiparticle field \cite{ginzburgbec}, our setup comprises a full-blown relativistic quantum field. In this way, we give an example of how  quantum technologies can help not only to expand the frontiers of our technical abilities but also to explore the frontiers of theoretical physics.

\begin{acknowledgments} Financial support from the Fundaci{\'o}n General CSIC is acknowledged by C.S. We also acknowledge support from Spanish MINECO/FEDER FIS2015-69983-P and FIS2015-70856-P, Ram\'on y Cajal Grant RYC-2012-11391, CAM PRICYT Project QUITEMAD+ S2013/ICE-2801, Basque Government IT986-16, and UPV/EHU UFI 11/55.
\end{acknowledgments}

\end{document}